\documentclass[twocolumn,english,prl, amsmath,amssymb,aps,superscriptaddress]{revtex4-1}
\usepackage[latin9]{inputenc}
\usepackage{amsmath}
\usepackage{graphicx}

\makeatletter
\usepackage{dcolumn}
\usepackage{bm}

\usepackage{array}
\usepackage{longtable}
\usepackage{textcomp}
\usepackage{multirow}
\usepackage{amsmath}
\usepackage{amssymb}
\usepackage{graphicx}
\usepackage{dcolumn}
\usepackage{bm}
\usepackage{setspace,lipsum}

\makeatletter

\pdfpageheight\paperheight
\pdfpagewidth\paperwidth

\providecommand{\tabularnewline}{\\}

\makeatother

\usepackage{babel}
\begin{document}

\preprint{APS/123-QED}

\title{Two-point active microrheology in a viscous medium exploiting a motional
resonance excited in dual-trap optical tweezers \\
}

\author{Shuvojit Paul$ $}

\affiliation{Indian Institute of Science Education and Research, Kolkata, India}

\author{Randhir Kumar}

\affiliation{Indian Institute of Science Education and Research, Kolkata, India}

\author{Ayan Banerjee}
\email{ayan@iiserkol.ac.in}

\affiliation{Indian Institute of Science Education and Research, Kolkata, India}

\date{\today}
\begin{abstract}
Two-point microrheology measurements from widely
separated colloidal particles approach the bulk viscosity of the host
medium more reliably than corresponding single point measurements. In addition, active microrheology offers the advantage of enhanced signal to noise over passive techniques. Recently, we reported the observation of a motional resonance induced in a probe particle in dual-trap optical tweezers when the control particle was driven externally [Paul et al. Phys. Rev. E {\bf 96}, 050102(R) (2017)]. We now demonstrate that the amplitude and phase characteristics of the motional resonance can be used as a sensitive tool for active two-point microrheology to measure the viscosity
of a viscous fluid. Thus, we measure the viscosity of viscous liquids 
from both the amplitude and phase response of the resonance, and demonstrate
that the zero-crossing of the phase response of the probe particle with respect to the external drive is superior compared to the amplitude response in measuring viscosity at large particle separations. We compare our viscosity measurements with that using a commercial rheometer and obtain an agreement $\sim1\%$. The method can be extended to viscoelastic material where the frequency dependence of the resonance may provide further accuracy for active microrheological measurements. 
\end{abstract}


\maketitle

\section{Introduction}

Microrheology measurements have enabled the measurement of viscosity using trace volumes of liquids and have proved to be especially useful in the context of biology, where the rheology of cellular environments naturally involve extremely small sample volumes \cite{frases09,robertson13,watts13}. Microrheology is performed using Brownian tracer probes, and can be performed  using both 'passive' and `active' methods. In the former, the thermal motion of the Brownian probes embedded and diffusing freely in the medium is measured \cite{hansen86,mason95,mason97,mackintosh99,wirtz09,squires10} while in the latter, an external force is applied to the probe(s) via optical or magnetic tweezers, and the response determined \cite{ouyang99,neuman08,mizuno08,brau07,chiang11,neckernuss16}. It is straightforward to comprehend that active microrheology provides greater signal-to-noise in measurements compared to the passive method, since thermal motion of the probes is manifested by the auto or cross-correlation amplitudes of their Brownian displacements which are often weak and therefore limited by experimental  noise. In contrast, active microrheology is performed by an external forcing of the probe with amplitudes much greater than that of the inherent Browninan flucturations, so that the corresponding response is also much larger \cite{brau07,mizuno08}. 

However, both active and passive microrheology measurements mostly use a single probe particle embedded in the medium. This is known as one-point microrheology, which has been implemented by both passive and active techniques, as mentioned earlier.  Optical traps - with their ability to provide long measurement times are ideal for microrheology applications so that they are also often used useful for passive microrheology \cite{brau07,mizuno08,watts13,tassieri15,tassieri10,tassieri12,scirepbayes}. In addition, they allow the possibility of detection over a large frequency bandwidth with the use of  position sensitive photodetectors that have much larger bandwidth compared to standard video microscopy that is typically preferred for freely diffusing probes. It must be noted, though, that measurements of the local viscosity using a single probe implies that these are over a very localized region of the sample, which may not necessarily represent the bulk viscosity very accurately. This may be due to the presence of inhomogeneities in temperature or density - the latter being particularly relevant for complex or viscoelastic material. In addition, determining rheological parameters from one-point microrheology by employing optical tweezers may be erroneous since the high intensity of the tweezers laser may lead to local heating and an enhanced temperature of the liquid in the immediate vicinity of the probe \citep{peterman2003laser}. Thus, two-point microrheology often provides more reliable estimation of the bulk viscosity/viscoelasticity as has been pointed out in the literature \cite{crocker2000two,crocker2007multiple,levine2000one,levine2001two,hormel2015two}. Two point passive microrheology using optical tweezers has also been performed \cite{starrs03}, but instances of active two-point microrheology using tweezers are rare. It is important to note that in the case of passive microrheology - the hydrodynamic interaction between two Brownian probes decreases inversely \citep{landau1986theory,doi1988theory} with their separation and hence it is challenging to experimentally measure such cross-correlations with large signal to noise when the inter-particle distance is large. Thus, it is preferable to develop optical tweezers-based active two-point microrheology techniques in order to improve the efficacy of such measurements.

In this paper, we describe a two-point active microrheology approach using
dual trap optical tweezers, where the trapped particles are widely
separated - thus yielding viscosity values that are very close to the bulk
viscosity of a viscous fluid. In our method, we drive one of the
trapped particles in the dual trap sinusoidally, and exploit the fact
that hydrodynamic interactions between the particles leads to a motional
resonance in the driven particle at a particular drive frequency.
Thus, we measure the resonance frequency from both the maximum amplitude
of the driven particle and the zero-crossing of its phase with respect
to the drive frequency. The value of this resonance frequency is dependent
on the medium viscosity - all other factors (such as trap stiffnesses
and particle separation) being unchanged - so that any shifts will
correspond to changes in viscosity of the fluid medium. We validate
the technique in water where we compare our measured values of viscosity
with the standard value for water. We also measure the accuracy of
the technique at different particle separations and find that the
the zero crossing of the phase provides more accurate results. We
then proceed to determine the viscosity of a water and glycerol solution
with different concentrations of glycerol. Our measured viscosity
values agree with that obtained using a commercial rheometer at the
level of $\sim1\%$. Our method is naturally extendable to viscoelastic or
even active material where the effect of retarded hydrodynamic interactions
as well the dispersion in viscosity could lead to more intriguing
characteristics of the resonance which may increase the accuracy as
well as sensitivity of active two-point microrheology. 

\section{Theoretical analysis\label{sec:Theoretical-analysis}}

The detailed theoretical treatment for the amplitude and phase response
of the driving (control) and driven (probe) particles in two hydrodynamically
coupled optical traps has been given in \citep{paul2017direct}. Here,
we provide a brief description for completeness. The Langevin equations
describing the stochastic trajectories of the trapped particles are
given by \citep{gardiner1985handbook}
\begin{align}
m_i{\bf \dot{v}}_{i} & =-\gamma_{ij}{\bf v}_{j}-{\bf \nabla}_{i}U+{\boldsymbol \xi}_{i}\label{eq1}\\
{\bf \dot{R}}_{i} & ={\bf v}_{i}
\end{align}
where $i,j=1,2$ refer to the driving and driven particle, $m_{i}$
are their masses, $\boldsymbol{v}_{i}$ are their velocities, $\boldsymbol{\gamma}_{ij}$
are the second-rank friction tensors encoding the velocity-dependent
dissipative forces mediated by the fluid, $U=U_{1}+U_{2}$ is the
total potential of the conservative forces, and $\boldsymbol{\xi}_{i}$,
the Langevin noises, are zero-mean Gaussian random variables whose
variance is provided by the fluctuation-dissipation relation$\langle\boldsymbol{\xi}_{i}(t)\boldsymbol{\xi}_{j}(t^{\prime})\rangle=2k_{B}T\boldsymbol{\gamma}_{ij}\delta(t-t^{\prime})$.
The bold-face notation, with Cartesian indices suppressed, is used
for both vectors and tensors in the above Eq. ${\bf \dot{R}}_{i}$
and ${\bf \dot{v}}_{i}$ are position and velocity of the i-th particle,
respectively. The optical potential is given by $U(t)=\frac{1}{2}\sum k_{i}({\bf R}_{i}-{\bf R}_{i}^{0})^{2}$
where ${\bf R}_{i}^{0}$ is the position of the potential minimum
of the i-th optical trap and $k_{i}$ is the corresponding stiffness.
In the experimental setup, the minimum of one of the optical trap
(driving particle) is shifted with an external periodic signal and
the response of the particle in the other trap (driven particle) is
studied. Assuming momentum to be rapidly relaxing on the time scale
of the trap motion, we neglect inertia and average over the noise
to get, 
\begin{align}
-\gamma_{ij}{\bf \dot{R}}_{j}-{\bf \nabla}_{i}U={\bf 0}\label{eq3}
\end{align}
Eq.$\ref{eq3}$ can be inverted and arranged to be represented in
terms of mobility matrices $\boldsymbol{\mu}$ in the following manner
(considering components of the matrix, where $\boldsymbol{\delta}$
is the $3\times3$ identity matrix), 
\begin{align}
{\bf \dot{R}}_{1} & =-\mu_{11}{\bf \mathbf{\boldsymbol{\delta}}}k_{1}({\bf R}_{1}-{\bf R}_{1}^{0})-{\boldsymbol{\mu_{\textrm{12}}}}k_{2}({\bf R}_{2}-{\bf R}_{2}^{0})\label{eq4}\\
{\bf \dot{R}}_{2} & =-{\boldsymbol{\mu_{\textrm{21}}}}k_{1}({\bf R}_{1}-{\bf R}_{1}^{0})-\mu_{22}{\bf \boldsymbol{\delta}}k_{2}({\bf R}_{2}-{\bf R}_{2}^{0})
\end{align}
Thus, incorporating the mobility matrices with separation vector
as the average distance between two the minima of the optical traps
we have, with ${\textstyle k_{i}\mathbf{R_{\textrm{i}}^{0}={\bf F}_{\textrm{i}}^{0}}}$,
\begin{align}
\frac{d}{dt}\begin{bmatrix}{\bf R}_{1}\\
{\bf R}_{2}
\end{bmatrix}=-\begin{bmatrix}\mu_{11}k_{1}{\bf \boldsymbol{\delta}} & {\boldsymbol{\mu_{\textrm{12}}}}k_{2}\\
{\boldsymbol{\mu_{\textrm{21}}}}k_{1} & \mu_{22}k_{2}{\bf \boldsymbol{\delta}}
\end{bmatrix}\begin{bmatrix}{\bf R}_{1}\\
{\bf R}_{2}
\end{bmatrix}+\begin{bmatrix}\mu_{11}{\bf \boldsymbol{\delta}} & {\boldsymbol{\mu_{\textrm{12}}}}\\
{\boldsymbol{\mu_{\textrm{21}}}} & \mu_{22}\boldsymbol{\delta}
\end{bmatrix}\begin{bmatrix}{\bf F}_{1}^{0}\\
{\bf F}_{2}^{0}
\end{bmatrix}\label{eq6}
\end{align}
The steady state solution of Eq.$\ref{eq4}$ can easily be calculated
by working in the frequency domain. Thus, assuming ${\bf A}=\begin{bmatrix}\mu_{11}k_{1}{\bf \boldsymbol{\delta}} & {\boldsymbol{\mu_{\textrm{12}}}}k_{2}\\
{\boldsymbol{\mu_{\textrm{21}}}}k_{1} & \mu_{22}k_{2}{\bf \boldsymbol{\delta}}
\end{bmatrix}$ and ${\bf M}=\begin{bmatrix}\mu_{11}{\bf \boldsymbol{\delta}} & {\boldsymbol{\mu_{\textrm{12}}}}\\
{\boldsymbol{\mu_{\textrm{21}}}} & \mu_{22}\boldsymbol{\delta}
\end{bmatrix}$, and taking Fourier transforms, we obtain, 
\begin{align}
{\bf R}_{i}(\omega)=\bigl[[-i\omega{\bf \boldsymbol{\delta}}+{\bf A}]^{-1}{\bf M}\bigr]_{ij}{\bf F}_{j}^{0}(\omega)={\boldsymbol{\chi}}_{ij}(\omega){\bf F}_{j}^{0}(\omega)\label{eq7}
\end{align}
where we now introduce the response ${\boldsymbol{\chi}}$, since one of the optical
traps is modulated by a sinusoidal signal of driving frequency $\Omega$,
which in the frequency domain may be represented as ${\bf F}_{j}^{0}(\omega)=\frac{X_{j}}{2}(\delta(\omega-\Omega)+\delta(\omega+\Omega))$,
with ${\displaystyle X_{j}}$ being the amplitude of the drive. Further,
${\boldsymbol{\chi}}$ is a block-diagonal matrix in cartesian indices. Given
the experimental set up, ${\boldsymbol{\chi}}$ can be decomposed into ${\boldsymbol{\chi}_{\parallel}}$
and ${\boldsymbol{\chi}_{\perp}}$ for motion along the trap separation and
the motion perpendicular to it. Inserting this form of the response
into Eq. \ref{eq7} and going back into the time domain, we have 
\begin{align}
R_{\parallel i}(t)=\chi'_{\parallel ij}\cos(\Omega t)X_{j}+\chi"_{\parallel ij}\sin(\Omega t)X_{j}\label{eq8}
\end{align}
Two time scales from two traps can be calculated to be $\tau_{i}=\frac{1}{\mu k_{i}}$,
so that,
\begin{align*}
{\boldsymbol{\chi}}_{\parallel}(\Omega)=\frac{1}{Det\boldsymbol{A}_{\parallel}-\Omega^{2}-i\Omega Tr\boldsymbol{A}_{\parallel}}\times\\
\begin{bmatrix}(\mu_{11}k_{1}-i\Omega) & \mu_{12}k_{2}\\
\mu_{21}k_{1} & (\mu_{22}k_{2}-i\Omega)
\end{bmatrix}^{-1}\begin{bmatrix}\mu_{11} & \mu_{12}\\
\mu_{21} & \mu_{22}
\end{bmatrix}\\
=\frac{1}{Det\boldsymbol{A}_{\parallel}-\Omega^{2}-i\Omega Tr\boldsymbol{A}_{\parallel}}\times\\
\begin{bmatrix}(\mu_{22}k_{2}-i\Omega) & -\mu_{12}k_{2}\\
-\mu_{21}k_{1} & (\mu_{11}k_{1}-i\Omega)
\end{bmatrix}\begin{bmatrix}\mu_{11} & \mu_{12}\\
\mu_{21} & \mu_{22}
\end{bmatrix}
\end{align*}
\begin{align}
{\boldsymbol{ \chi}}_{\parallel}(\Omega)=\frac{Det\boldsymbol{A}_{\parallel}-\Omega^{2}+i\Omega Tr\boldsymbol{A}_{\parallel}}{(Det\boldsymbol{A}_{\parallel}-\Omega^{2})^{2}+\Omega^{2}(Tr\boldsymbol{A}_{\parallel})^{2}}\times\nonumber \\
\begin{bmatrix}k_{2}Det\boldsymbol{M}_{\parallel}-i\mu\Omega & -i\Omega\mu_{12}\\
-i\Omega\mu_{21} & k_{1}Det\boldsymbol{M}_{\parallel}-i\mu\Omega
\end{bmatrix}\label{eq:9}
\end{align}
We are interested in the resonance in amplitude of the probe due to
the forcing of the control, which is obtained by maximizing the modulus
of ${\boldsymbol {\chi_{\parallel}}}_{21}$ with respect to $\Omega$. Thus, 
\begin{align}
\mid{\boldsymbol {\chi_{\parallel}}}_{21}\mid & =\mid\frac{-i\Omega\mu_{21}}{Det\boldsymbol{A}_{\parallel}-\Omega^{2}-i\Omega Tr\boldsymbol{A}_{\parallel}}\mid\nonumber \\
 & =\frac{\Omega\mu_{21}}{\sqrt{(Det\boldsymbol{A}_{\parallel}-\Omega^{2})^{2}+\Omega^{2}(Tr\boldsymbol{A}_{\parallel})^{2}}}\label{eq:10}
\end{align}
and finally, the amplitude of the probe is given by 
\begin{align}
A=\mid{\boldsymbol {\chi_{\parallel}}}_{21}\mid k_{1}X & =\frac{\Omega X\mu_{21}k_{1}X}{\sqrt{(Det\boldsymbol{A}_{\parallel}-\Omega^{2})^{2}+\Omega^{2}(Tr\boldsymbol{A}_{\parallel})^{2}}}
\end{align}
, where $X$ is the amplitude of the driving signal on the control
particle. The phase response of the probe with respect to the drive
is given by 
\begin{align}
\Delta\phi & =\tan^{-1}\left(\frac{\Omega^{2}-Det\boldsymbol{A}_{\parallel}}{\Omega(Tr\boldsymbol{A}_{\parallel})}\right)\label{eq11}
\end{align}
Clearly the resonance frequency in dimensionless unit is $\tau_{1}\Omega_{res}=\sqrt{\dfrac{Det\boldsymbol{A}_{\parallel}}{\mu_{11}^{2}k_{1}^{2}}}=\sqrt{\dfrac{k_{2}}{k_{1}}\left(1-\dfrac{\mu_{12}^{2}}
{\mu_{11}^{2}}\right)}$
when $\mu_{12}\neq0$. 

Note that Eq. \ref{eq:10} is analogous to the expression for displacement
of a forced harmonic oscillator, with the resonance frequency in this
case given by $\omega_{0}^{2}=Det\boldsymbol{A}_{\parallel}$ and the damping term
$\Gamma=Tr\boldsymbol{A}_{\parallel}$. Finally, we have
\begin{align}
\omega_{0}^{2}=Det\boldsymbol{A}_{\parallel}=k_{1}k_{2}\left(\mu_{11}\mu_{22}-\mu_{12}\mu_{21}\right)\label{eq12}
\end{align}
\begin{align}
\Gamma=Tr\boldsymbol{A}_{\parallel}=\left(k_{1}\mu_{11}+k_{2}\mu_{22}\right)\label{eq13}
\end{align}
with $\mu_{ij}$ is the component of  $\boldsymbol{\mu}_{ij}$ that is parallel to ${{\bf (R_1^0 - R_2^0)}}$. Here, $\mu_{11}=\mu_{22}=\frac{1}{6\pi\eta a}$, $\mu_{12}=\mu_{21}=\frac{1}{8\pi\eta r_{0}}\left(2-\frac{4a_{0}^{2}}{3r_{0}^{2}}\right)$, $a_{0}$
is the radius of the spherical particle, and $r_{0}$ is the trap
separation.
\begin{figure}
\centering \includegraphics[scale=0.32]{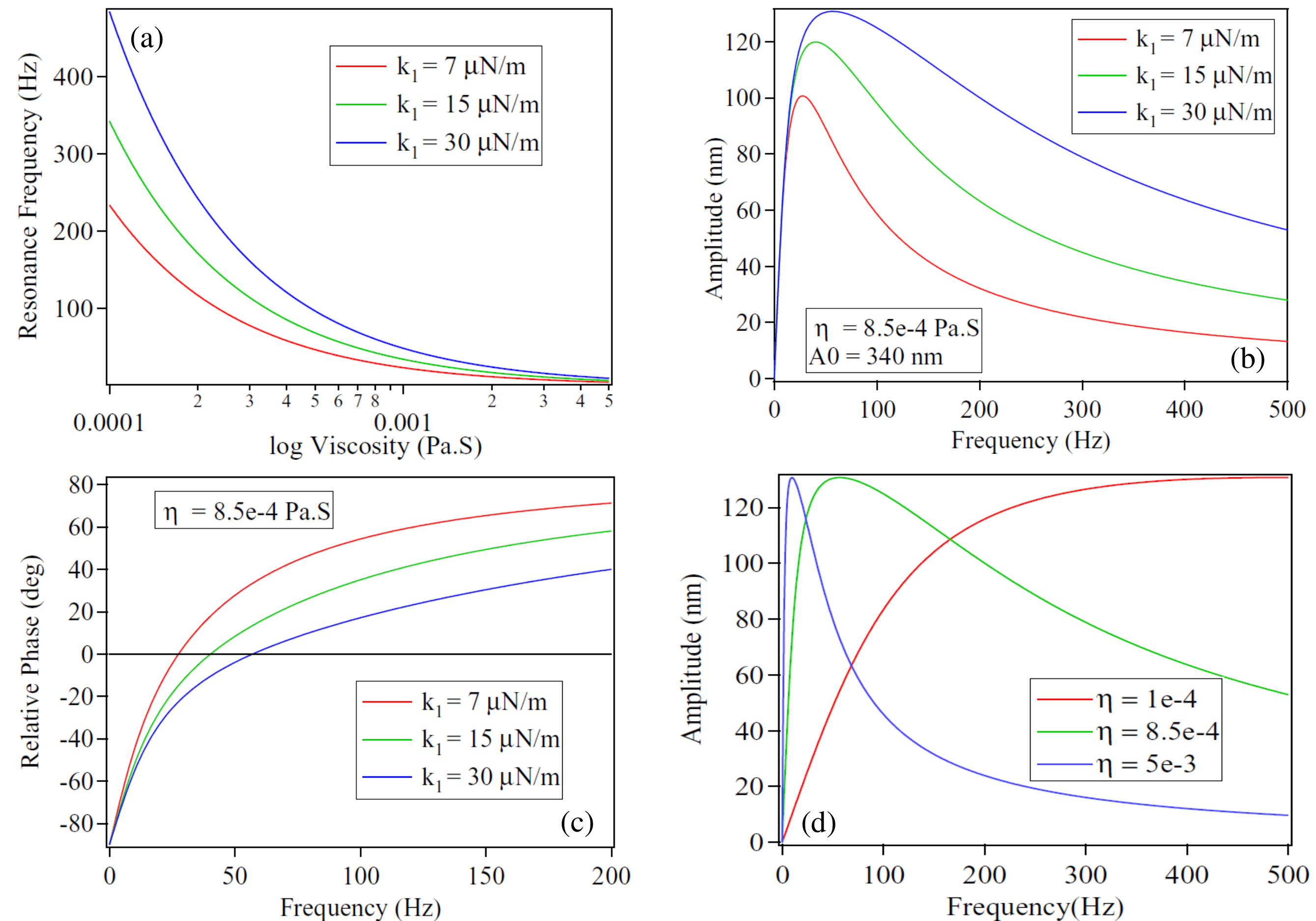}
\caption{(a) Resonance frequency as a function of viscosity for different stiffnesses. $\mathit{k_{2}}$ is kept constant (at 3 ${\textstyle {\displaystyle \mu N/m}}$), while $\mathit{k_{1}}$ is varied - the red, green, and blue lines
signifying $\mathit{k_{1}}$ values of 7, 15, and 30 ${\textstyle {\displaystyle \mu N/m}}$, respectively. (b) Amplitude of probe as a function of drive frequency for same stiffness values and colour code as (a). (c) Phase of probe
as a function of drive frequency. (d) Amplitude of probe as a function
of drive frequency for stiffness $k_{2}$= 3 ${\textstyle {\displaystyle \mu N/m}}$, $k_{1}$= 7 ${\textstyle {\displaystyle \mu N/m}}$ and different viscosity ($\eta$) values of 1e-4 (red), 8.5 e-4 (green), and 5 e-3
(blue).}
\label{sensitivity} 
\end{figure}

An inspection of Eq. \ref{eq12} reveals that $\omega_{0}$ is dependent
on the stiffness $k_{1}$, $k_{2}$ and of the viscosity $\eta$ of
the medium, which implies that for constant stiffness, any change
in $\eta$ would be revealed in shifts in $\omega_{0}$. We now perform
a set of theoretical simulations to find out the implications of these
equations in the determination of viscosity of the ambient liquid
where the particles are embedded. First, we study the change in $\omega_{0}$
with changing $\eta$ for different stiffness ratios of $k_{1}$ and
$k_{2}$. Thus, we keep the value of $k_{2}$ constant ($k_{2}=3\:\mu N/m$,
similar to what we use in experiments as described later), and use
different values of ($k_{1}=7,\:15,\:30\:\mu N/m$) to find out how
$\omega_{0}$ varies with changing $\eta$ over a range 1 e-4 - 5
e-3 Pa-s. We choose this range to mimic experimentally feasible conditions,
noting that optical trapping is rendered rather difficult in liquids
with viscosity lesser or larger than water by around an order of magnitude
(viscosity of water being 8.5e-4 Pa-s at 300K). This is shown in Fig.
\ref{sensitivity}(a). It is clear that $\omega_{0}$ changes the
most for the largest ratio of trap stiffnesses, i.e. when the product
is largest (as is clear from Eq.\ref{eq12}). However, it is incorrect
to conclude that the experimental sensitivity in measuring $\eta$
is also highest for the highest stiffness ratio. Note that the measurement
sensitivity would not only be dependent on shift in $\omega_{0}$,
but also on how sensitively the shift can be measured from the amplitude
and phase characteristics of the resonance. This essentially translates
to the width of the resonance in terms of amplitude, and the change
in phase across the resonance frequency. Thus, we study the the amplitude
and phase response of the resonance for the same ratio of stiffnesses,
as shown in Fig. \ref{sensitivity}(b) and (c). It is very clear that
the width of the resonance increases with increasing stiffness ratio,
so that the narrowest width is obtained for $k_{1}=7\:\mu N/m$ for
$k_{2}=3\:\mu N/m$ (Fig. \ref{sensitivity}(b)). The maximum change
of phase also occurs for the same combination of stiffness as shown
in Fig. \ref{sensitivity}(c). Interestingly, the resonance peak-width
reduces with increasing viscosity, as we observe in Fig. \ref{sensitivity}(d),
where we plot the amplitude for three different viscosities ($\eta$=
1, 8.5, and 50 e-4 Pa.s) as shown in the figure. It is therefore clear
that we have several different possibly competing effects that would
finally decide the sensitivity for viscosity measurement. In order
to obtain a quantitative understanding of this, we combine the shifts
in frequency ($\Delta\omega_{0}$) and that in resonance peak-width/phase
change so as to write down the following relations for the phase ($\Phi_{s}$)
and amplitude sensitivities ($A_{s}$) as:
\begin{align}
\Delta\omega_{0}=\frac{\partial\omega_{0}}{\partial\eta}\Delta\eta=
\frac{\eta\omega_{0}\Delta\eta}{\eta^{2}-\left(\frac{\Delta\eta}{2}\right)^{2}}\label{eq14}
\end{align}
\begin{align}
\Phi_{s}=\Phi(\omega_{0}+\Delta\omega_{0})-\Phi(\omega_{0})=
\tan^{-1}\left(\frac{\Delta\omega_{0}\left(2\omega_{0}+\Delta\omega_{0}\right)}{\Gamma\left(\omega+\Delta\omega_{0}\right)}\right)\label{eq15}
\end{align}
\begin{align}
A_{s}=A(\omega_{0}+\Delta\omega_{0})-A(\omega_{0})=
\nonumber \end{align}
\begin{equation}
\frac{\mu_{21}k_{1}X\left(\omega_{0}+\Delta\omega_{0}\right)}{\sqrt{\left(\left(\omega_{0}
+\Delta\omega_{0}\right)^{2}-\omega_{0}^{2}\right)^{2}+\left(\left(\omega_{0}+\Delta\omega_{0}
\right)\Gamma\right)^{2}}}-\frac{\mu_{21}k_{1}X}{\Gamma}\label{eq16}
\end{equation}
where , $\Delta\omega_{0}$ is the shift in resonance frequency due
to $\Delta\eta$ change in viscosity around the viscosity $\eta$.
$\Phi_{s}$ and $A_{s}$ are defined as the phase and amplitude sensitivity,
respectively, and represent the changes in amplitude and phase due
to change in viscosity $\Delta\eta$ at a particular resonance frequency,
keeping all other parameters same.

\begin{widetext}
\begin{figure*}
\includegraphics[scale=0.45]{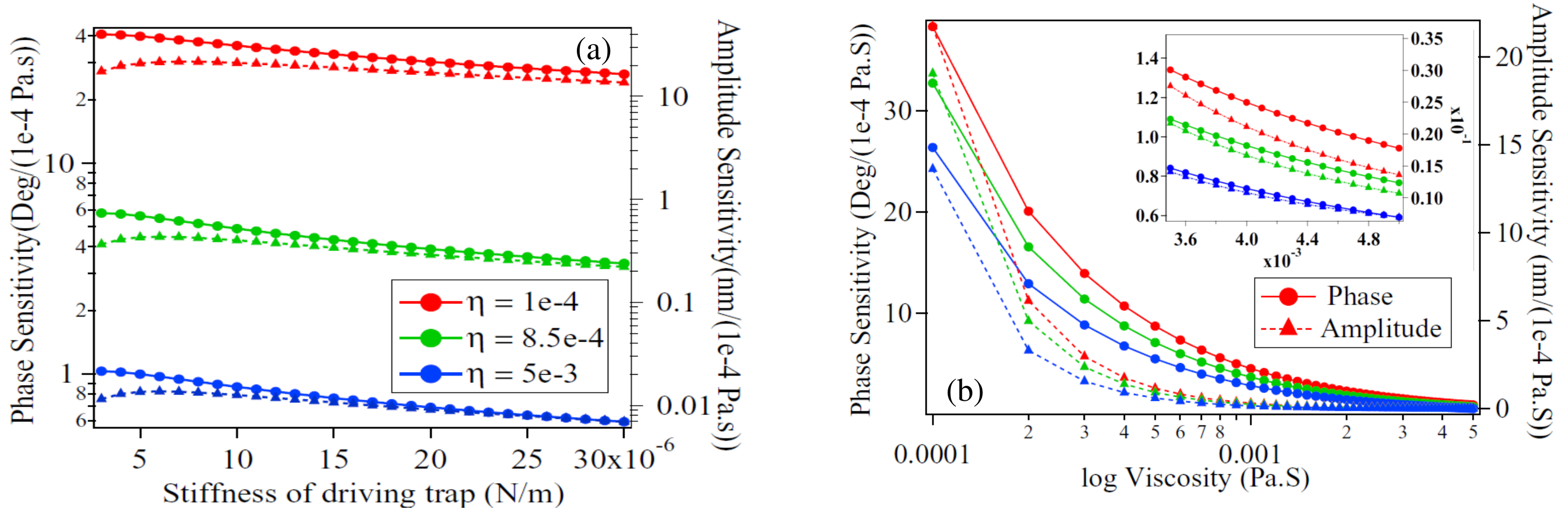}\caption{Phase (left axis, filled solid circles) and amplitude (right axis,
filled solid triangles) sensitivities as a function of (a) $k_{1}$ and
(b) (log) viscosity. Once again, $\mathit{k_{2}}$ is kept constant (at 3 ${\textstyle {\displaystyle \mu N/m}}$), while $\mathit{k_{1}}$ is varied - the red, green, and blue lines signifying $\mathit{k_{1}}$ values of 7, 15, and 30 ${\textstyle {\displaystyle \mu N/m}}$, respectively. In the inset, a zoomed-in version of the phase and amplitude sensitivities for large viscosity values is shown.}
\label{viscositysensitivity}
\end{figure*}
\end{widetext}

It is now necessary to study $A_{s}$and $\Phi_{s}$ carefully in
order to obtain a complete understanding of the efficacy of our technique
in measuring viscosity. Thus, we first proceed to plot $A_{s}$ and
$\Phi_{s}$ against increasing $k_{1}$, keeping $k_{2}$ fixed at
$3\:\mu N/m$, at 3 different fixed values of viscosity of the surrounding
liquid (same as Fig. \ref{viscositysensitivity}(b)). This is shown
in the right axis of the plot depicted in Fig.\ref{viscositysensitivity}(a).
We observe that $A_{s}$ does not change substantially when we vary
the stiffness of the driving trap ($k_{1}$), so that it is clear
that the increase in resonance frequency shift and the corresponding
increase in peak-width almost compensate each other. For $A_{s}$,
we have a maximum sensitivity around $k_{1}=7\:\mu N/m$. However,
the change is only about 25\% in the entire range of $k_{1}$. In
addition, the sensitivity reduces monotonically for increasing viscosity,
so that we have the lowest sensitivity for the highest viscosity.
This behaviour is also seen in $\Phi_{s}$ (left axis of the same figure).
We notice that $\Phi_{s}$, unlike $A_{s}$, decreases linearly with
increasing $k_{1}$, but the change is also only around 35\% over
the entire range. It is also clear that the value of the phase sensitivity
is higher than the amplitude (tens of degrees, as compared to few
nm), which makes the phase measurement less challenging for experiments. 
We now consider a situation where we continuously change the value
of viscosity of the surrounding liquid, something which may arise
in a practical situation for online monitoring of the viscosity of
a liquid\textit{ in situ.} We use the three different stiffness ratios
we had used earlier in Figs. \ref{sensitivity}(a)-(c). Once again,
the phase sensitivity is indeed higher than the amplitude sensitivity.
Following the trend observed in Fig. \ref{viscositysensitivity}(a),
both $A_{s}$ and $\Phi_{s}$ are clearly dependent on the viscosity
value being measured, with the sensitivity decreasing with increasing
viscosity, though $\Phi_{s}$ changes less drastically than $A_{s}$.
We attempt to provide an explanation for this. From Eq. \ref{eq13},
we observe that the damping $\Gamma$ depends on the mobility, which
obviously reduces with increasing viscosity so that the width of the
resonance actually reduces with increasing viscosity as we have demonstrated
in Fig. \ref{sensitivity}(d). However, it is clear from Eq. \ref{eq14},
that the resonance frequency changes as $\dfrac{1}{\eta^{2}},$ while
the damping changes as $\dfrac{1}{\eta}$. Thus, for large value of
$\eta,$ the narrowing of the width is accompanied by an even smaller
change of resonance frequency for a given change in $\eta$. The resultant
sensitivity of the resonance response - a combination of both factors -
is therefore dominated by the small change in the resonance frequency,
which makes the sensitivity lower for increasing viscosity. The change
in both phase and amplitude sensitivity also appear almost linear
for large viscosities, as shown in the inset of Fig. \ref{viscositysensitivity}(b).
It is also clear that both phase and amplitude sensitivities are highest
for the lowest stiffness ratio, which means that they have greater
dependence on the sharpness of the resonance than the shift in the
central frequency itself. Thus one can, in principle, get higher sensitivities
for higher viscosity values by choosing lower values of $k_{1}$ and
$k_{2}$, but these imply rather weak optical traps. Thus, we need
to perform a trade-off between having the best sensitivity and stable
optical traps, so that for our experiments we typically choose values
of $k_{2}=3\:\mu N/m$ and $k_{1}=7\:\mu N/m$, which result in a
viscosity sensitivity of around 0.43 nm/1e-4 Pa.s for $A_{s}$ and
5.3 deg/1e-4 Pa.s for $\Phi_{s}$ around the viscosity value of water.
Note that these were the stiffness values where we obtained largest
$A_{s}$ in Fig. \ref{viscositysensitivity}(a) . We also keep $k_{2}<k_{1}$
for two reasons: 1) To increase the amplitude detection sensitivity
of the probe, and 2) to minimize the effect of the probe on the control,
i.e. to reduce hydrodynamic back-flow effects. 

We now present our experimental setup and results of the measurements of viscosity of different viscous liquids.

\section{Experimental setup}
\begin{figure}
\centering \includegraphics[width=0.45\textwidth]{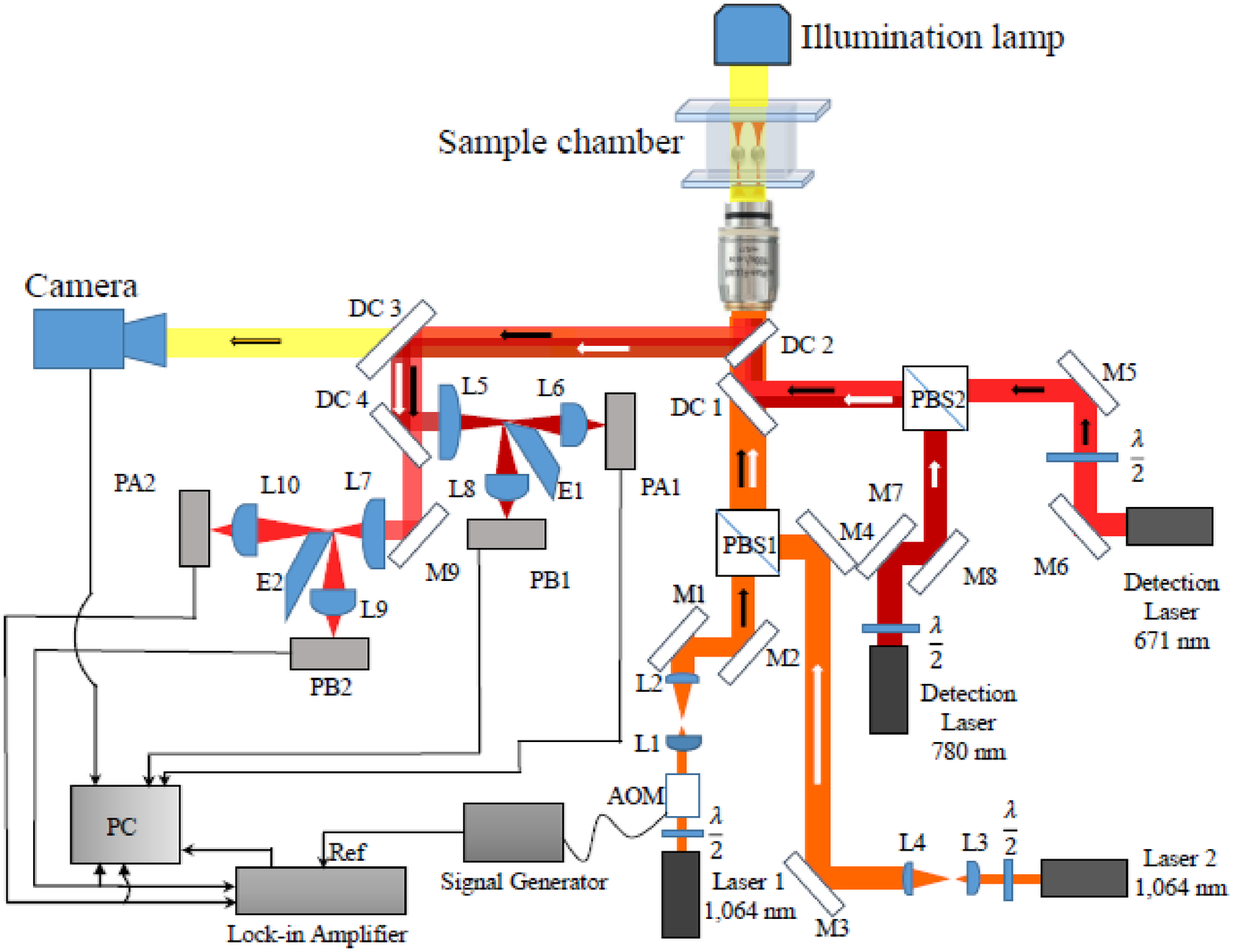}
\caption{Schematic of the experimental setup. $\frac{\lambda}{2}$: half wave
plate, L: lens, M: mirror, AOM: accousto optical modulator, PBS: polarizing
beam splitter, DC: dichroic mirror, E: edge mirror, P: polariser PA,
PB: photodiodes (Thorlabs PDA100A-EC), PC: Personal Computer.}
\label{Setup} 
\end{figure}
\begin{figure}
\includegraphics[width=0.3\textwidth]{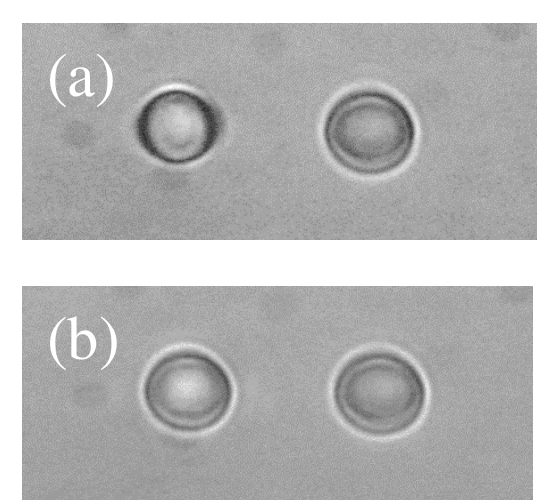}
\caption{ (a) The control and probe particles trapped in different z-planes,
which is apparent from the different sizes. (b) The particles trapped
in the same plane.}
\label{bead_height} 
\end{figure}

We used a high NA immersion-oil microscope objective (Zeiss PlanApo
100 x 1.4) to focus two orthogonally polarized laser beams of wavelength
$\lambda=1064$ nm to create a dual beam optical tweezers (fig : \ref{Setup}).
We employed two different $\frac{\lambda}{2}$ plates to render the
polarization state of the two laser beams orthogonal to each other,
and an acousto-optic modulator (AOM) to sinusoidally drive one of
the traps (driving particle). We used the lens pair L1-L2 to set up
a telescopic arrangement to ensure that the AOM was placed at a location
that was conjugate to the back focal plane of the microscope objective
to exactly project the angular deflection due to the AOM at the back
focal plane of the objective lens. To keep the intensity constant
(within 1$\%$), we scanned the first order diffracted beam from the
AOM by a very small amount resulting in a very small angular deflection
that was magnified at the trap focal plane by employing a long optical
path to the microscope objective. As demanded by our measurement scheme,
we maintained the other trap stationary. Note that the trap positions
along the z-axis can be controlled using the lens combination L3 and
L4. For our experiment to yield correct results, it is essential that
we trap the two particles in the same z-plane. Our control over particle
positions along the z-axis is shown in Figure \ref{bead_height}(a)
and (b). In (a), the particles are not in the same z-plane, as a result
of which they appear to be of different size on the image obtained
by the microscope camera. In (b), the images are identical, which
shows two positions of the two particles on the same z-axis. We prepared
two periscope systems by two sets of mirrors (M1, M2 and M3, M4) to
independently steer each beam in the x-y plane. Later, we used a polarizing
beam splitter (PBS1) to recombine the two trapping beams. The detection
lasers were of wavelengths $\lambda=671$ and $\lambda=780$ nm, and
were also prepared to have orthogonal polarization to each other.
They were combined by the polarizing beam splitter PBS2. The detection
laser beams were overlapped with the trapping laser beams by a dichroic
beam splitter (DC1). To detect the positions of the trapped particles
by the back-focal-plane-interferometry technique, and to image them
using white light from the microscope illumination lamp, the back
scattered light of the detection lasers and the white light were reflected
by the dichroic beam-splitter DC2 towards the detection system and
camera, respectively, after being filtered by a third dichroic beam-splitter
DC3. The back-scattered detection beams (of different wavelengths
as mentioned earlier) were finally separated by the dichroic beam-splitter
DC4. The separated detection beams were then focused by two lenses
L5 and L7, and balanced detection systems were set up by dividing
each beam half-way by edge mirrors E1 and E2, respectively, and focusing
them using lenses L6, L8, L10, L9 on photo diode pairs PA1 + PB1 and
PA2 + PB2, respectively. We prepared a sealed sample chamber by attaching
two microscope cover slips by double-sided tape so that the dimensions
were $20\times10\times0.2$ mm. We used samples of very low volume
fraction ($\phi\approx0.01$) with 3 $\mu$m diameter polystyrene
spherical particles suspended in a host liquid (water or water + glyercol)
constituting the sample. We trapped two such polystyrene particles
in two calibrated optical traps (dual optical tweezers) separated
by distances varying from 4.5 to 9 $\mu$m, and situated approximately
30 $\mu$m away from the nearest wall. We generated sinusoidal voltage
waveforms from a signal generator (Tektronix, AFG3022B) which were
coupled to the AOM to modulate the driving trap. The same we also
used as the reference signal of a commercial lock-in amplifier (Stanford
Research, SR830). We input the detection signals from the balanced
detector system for the control particle into the lock-in amplifier
to measure amplitude and phase response of the driven particle, and
into a computer to calibrate the trap. We used a low pass filter time
constant of around 2 m in the lock-in for each measurement. For the
viscosity measurements of the water and glycerol solutions, we prepared
different v/v mixtures of glycerol and water and measured the corresponding
viscosities in a commercial rheometer (Brookfield DB3TLVCJ0) to compare
with the measurements obtained using our method.

\section{Experimental results and discussions}
\label{sec:DAR}
\begin{figure}
\centering \includegraphics[width=0.45\textwidth]{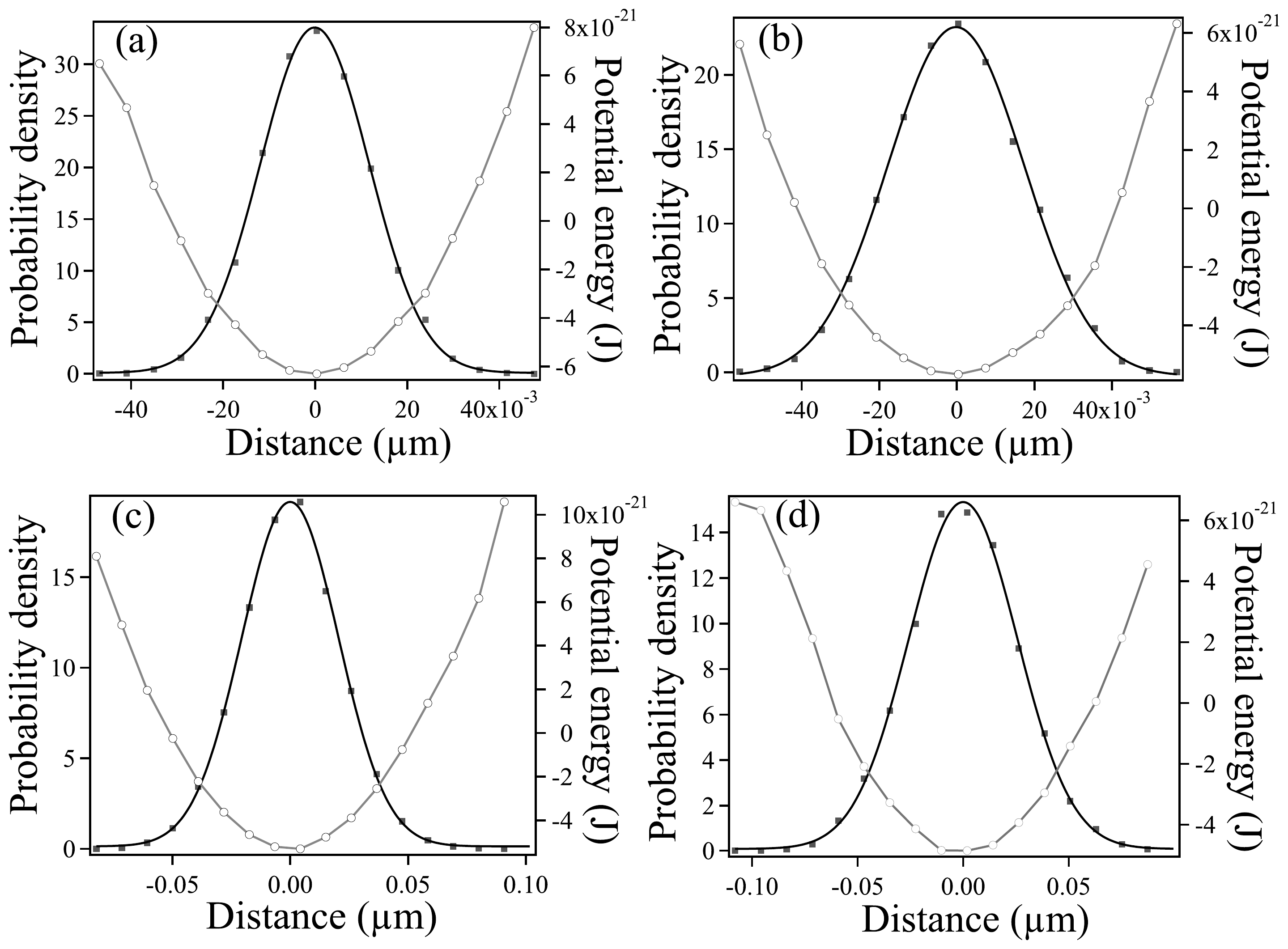}
\caption{(a) and (c) show the position probability density function plotted
against particle displacement for the control particle for trap separations
of 9 and 4 $\mu m$, respectively, while (b) and (d) show that for
the probe particle for the same separations. The solid circles in
black are the probability values computed from experimentally measured
data, which are then fit Gaussian functions (solid black lines). Each
figure also shows the corresponding trap potentials calculated from
the same experimental data (right axis, open grey circles), which
are fit to quadratic functions of the displacement (solid gray lines).}
\label{histogram} 
\end{figure}
To calibrate the individual optical traps independently, we recorded
the thermal fluctuations of both driving/driven particles individually
with the adjacent trap empty (the corresponding trapping beam was,
however, still on). We used the equipartition theorem to measure the
stiffness of the trapped single particle. It is important to note
that the equipartition method of trap calibration is independent of
the rheological nature of the medium. However, careful sensitivity
measurements of the detection systems are required to calibrate our
system. This we performed by shifting the center of the trap potential
by a known amount using the AOM and by measuring the corresponding
change in the signal from the balanced detection system. Since the
potential experienced by the trapped particle is equivalent to that
by a harmonic oscillator, the position probability need necessarily
be Gaussian, as demanded by the solution of the steady state Fokker-Planck
equation $\mathcal{N}(0,\frac{k_{B}T}{k})$. This probability distribution
thus reflects the harmonic nature of the trap which should be independent
of the separation between the two trapped particles. This entails
that the trapping potentials should remain Gaussian throughout our
experiment. In Fig. \ref{histogram}, we demonstrate that the potential
experienced by both the control and probe particles remain Gaussian
for the smallest and largest trap separations (4.5 and 9 $\mu m$,
respectively). This conclusively demonstrates that there exists no
optical cross talk between the adjacent traps. 
\begin{figure}
\centering \includegraphics[width=0.4\textwidth]{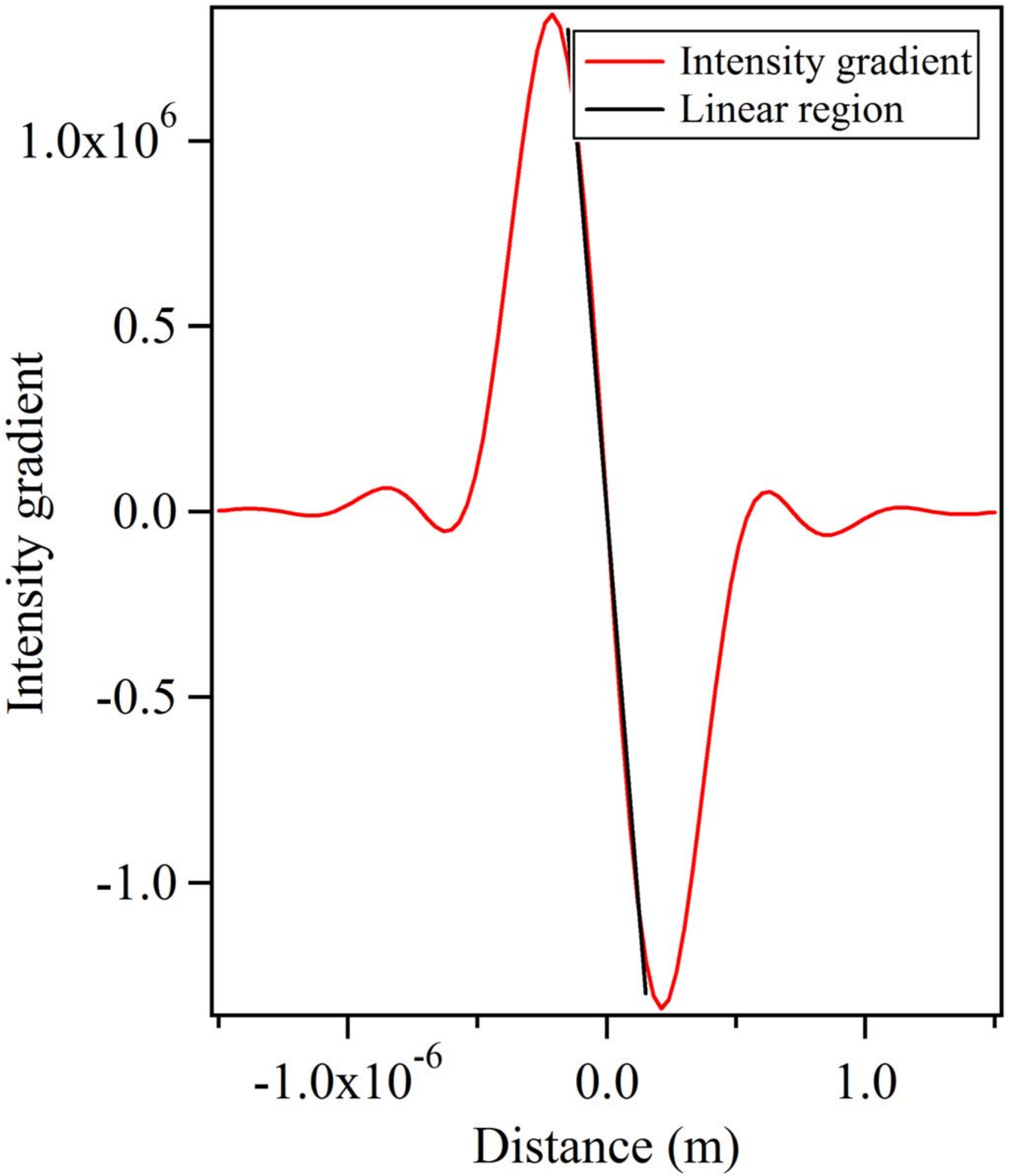}
\caption{Simulated intensity gradient using our trap parameters. The force
acting on the trapped particle is proportional to the intensity gradient
- the linear region of which appears to be about 0.3 $\mu m$ from
a straight line fit around the center.}
\label{linearity} 
\end{figure}
\begin{table}[b]
\caption{Measured viscosity of water using phase ($\eta_{p}$) and amplitude
($\eta_{a}$) for different particle separations.}
\begin{tabular}{|c|c|c|}
\hline 
Separation($\text{\textmu m}$)  & \multicolumn{1}{c|}{\ensuremath{\eta}$_{p}$(Pa.s)} & \ensuremath{\eta}$_{a}$(Pa-s)\tabularnewline
\hline 
4.5 & 0.00086$~\pm~4e-05$ & 0.0009$~\pm~2e-04$\tabularnewline
\hline 
7.0 & 0.0009$~\pm~5e-05$ & 0.0011$~\pm~1e-04$\tabularnewline
\hline 
9.0 & 0.00085$~\pm~2.5e-05$ & 0.00074$~\pm~1e-04$\tabularnewline
\hline 
\end{tabular}\label{tab:separation} 
\end{table}

For viscosity measurements on the dual-trap system, we first calibrated
each trap individually, and proceeded to measure the response of the
driven particle under the influence of the driving particle. As mentioned
in Section \ref{sec:Theoretical-analysis}, we chose the driving trap
stiffness greater than that of the driven trap (actual values used
are provided in Table \ref{tab:viscosity_table}). The oscillation
amplitude of the control was very small (0.25 $\mu m$ peak to peak)
compared to the actual separation between the two particles, we ensured
that the total particle amplitude was maintained within the linear
region of the trap. To find out an approximate value of the linear
region for our trap, we performed a simple simulation of the potential
given our trapping beam parameters as shown in Fig. \ref{linearity},
where we plot the field gradient against distance, and a linear fit
near the center shows that the extent of the linear region is around
0.3 $\mu m$, which is larger than the driving beam amplitude.
As mentioned earlier, the detection signal from the probe was fed
into a lock-in amplifier and the amplitude and phase with respect
to the drive were measured at different drive frequencies. Thus, we
determined both the zero-crossing of the phase as well as the amplitude
maximum of the probe as a function of the drive frequency. While we
perform both phase and amplitude measurements, it is obvious that
the phase measurement is more reliable, since obtaining sub-nm detection
sensitivities - as would be needed for measuring viscosity from the
amplitude shift - is indeed rather challenging.
\begin{figure}
\centering \includegraphics[scale=0.35]{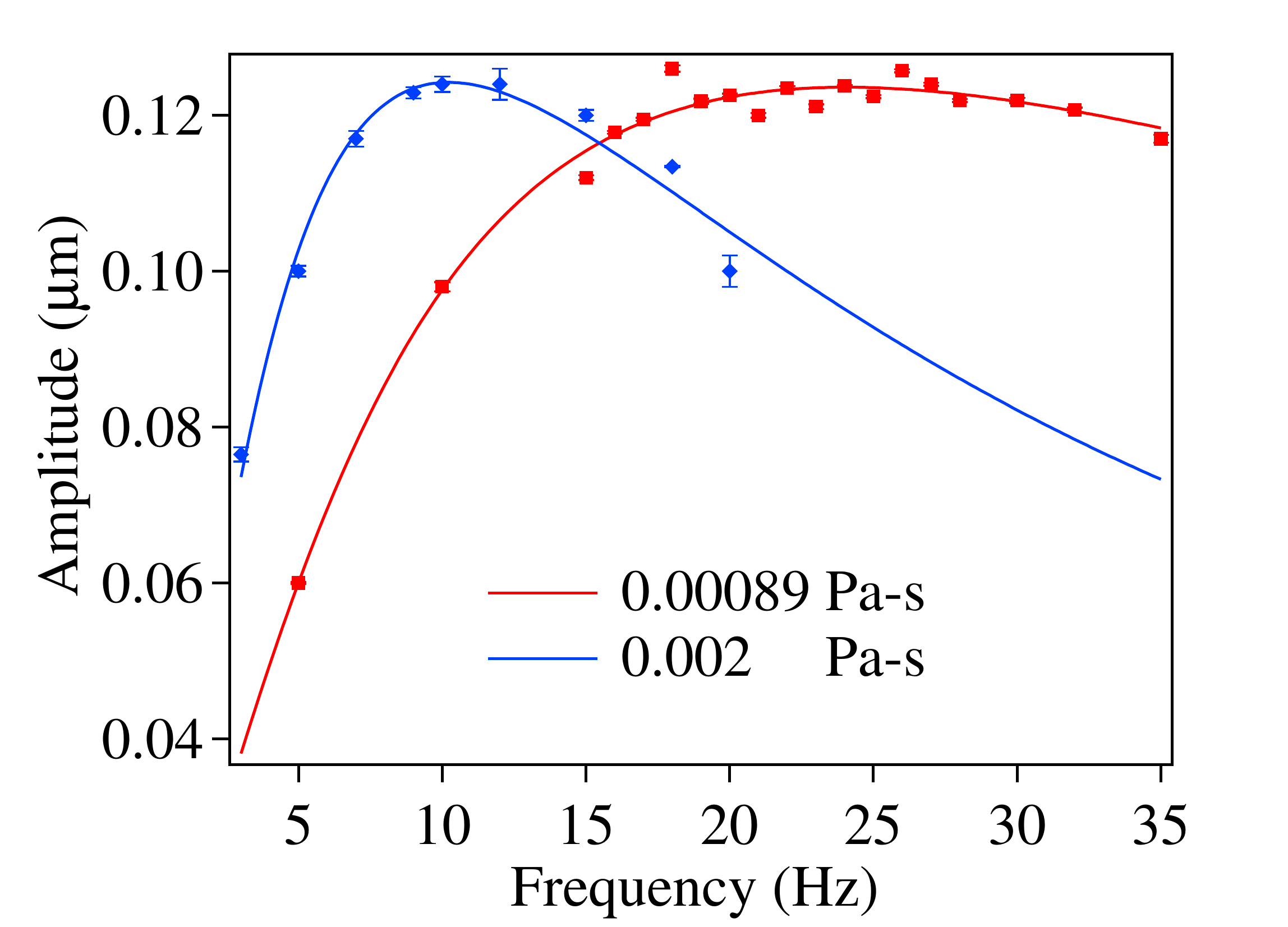}
\caption{Experimentally measured amplitude responses of the driven particle
for two different viscosity solutions. The experimentally measured
data are shown in red (blue) solid circles for viscosities of 0.00089
(red) and 0.002 (blue). The data is fit to the amplitude response
given in Eq.\ref{eq:10}.}
\label{viscosity figure} 
\end{figure}

Our next step was to determine the relative sensitivity of the amplitude
and phase measurement approaches. A reliable method towards achieving
this is to measure the respective responses as a function of particle
separation, since the hydrodynamic coupling reduces with increasing
distance. We performed measurements at three different particle separations:
4.5, 7, and 9 $\mu$m. Our aim was to determine the viscosity of water
from both amplitude and phase responses, since the standard value
of viscosity of water is very well-known (8.5 e-4 Pa-s at 300K). Thus,
we found that the zero-crossing of the phase is more accurate in determining
the viscosity reliably, as is shown in Table \ref{tab:separation}.
The value of viscosity of water as obtained from the zero-crossing
of the phase is both more accurate (within 2\%) and precise (statistical
error between 3-7\%) compared to the amplitude response - both corresponding
errors being almost double that compared to the phase response. The
amplitude response also becomes clearly less reliable as the particle
separation is increased, which is due to the fact that the signal
to noise of the measured data approaches the dark noise levels of
the detector as the hydrodynamic coupling reduces.

Finally, we measured the viscosity of different water + glycerol solutions
from the phase response of the resonance frequency for each solution.
The results are shown in Table \ref{tab:viscosity_table}, where $\eta$
is the viscosity measured in the commerical rheometer, and $\eta^{*}$
is the viscosity measured using our resonance technique. The trap
stiffnesses were kept the same for all measurements, and the corresponding
resonance frequencies ($\omega_{0}/2\pi$) shifted as shown in Table
\ref{tab:viscosity_table}. The amplitude response of the resonance
for the highest and lowest viscosity solutions are demonstrated in
Fig. \ref{viscosity figure}. The value of the resonance frequencies
are determined at 3-5\% precision, and the centers shift towards lower
frequency with increasing viscosity as predicted by the theory. The
measured values of viscosity are all within 2\% of the rheometer measured
values. We also note that the accuracy of the measured values we report
now have been improved by around 2 times over the measurements we
recently reported in Ref. \citep{scirepbayes}, where we performed passive microrheology by using a Bayesian analysis of the position autocorrelation of a single trapped Brownian particle to determine the viscosity of an unknown solution. It needs
to be mentioned here that both accuracy and precision can indeed be
improved by using even low stiffness ratios, which would yield sharper
resonances and correspondingly larger amplitude and phase sensitivities. 
\begin{widetext}
\begin{table*}
\centering{}\centering\caption{Measured viscosity ($\eta^{*}$) compared to that measured in rheometer ($\eta$) . The corresponding trap stiffnesses are also shown.}
\begin{spacing}{2}
\begin{longtable}{|c|c|c|c|c|}
\hline 
\multirow{2}{*}{$\eta$ (Pa.S) } & \multicolumn{2}{c|}{Stiffness (N/m) } & Resonance frequency (Hz)  & \multirow{2}{*}{$\eta^{*}$(Pa.S) }\tabularnewline
\cline{2-4} 
 & $k_{1}$  & $k_{2}$  & $\omega_{0}/2\pi$  & \tabularnewline
\hline 
0.00089  & 6.9(1)e-06  & 2.7(1)e-06  & $24.0\ \pm1.4$  & 0.00088(5) \tabularnewline
\hline 
0.00137  & 6.9(2)e-06  & 2.7(1)e-06  & $15.4\pm0.6$  & 0.00138(5) \tabularnewline
\hline 
0.0015  & 6.9(1)e-06  & 2.7(2)e-06  & $14.0\pm0.4$  & 0.00151(4) \tabularnewline
\hline 
0.001827  & 6.8(1)e-06  & 2.8(1)e-06  & $11.5\pm0.6$  & 0.00184(9) \tabularnewline
\hline 
0.0020 & 6.9(3)e-06  & 2.8(2)e-06  & $10.2\pm0.4$  & 0.00204(5) \tabularnewline
\hline 
\end{longtable}\label{tab:viscosity_table} 
\end{spacing}
\end{table*}
\end{widetext}

\section{Conclusions}

In conclusion, we demonstrate a two-point active rheology technique based on dual trap optical tweezers, where we measure the response of a stationary (probe) colloidal particle when the control particle is driven sinusoidally. The hydrodynamic interactions between control and probe lead to a motional resonance of the probe at a particular driving frequency of the control particle. The resonance frequency
is dependent on the stiffnesses of the two traps and the viscosity of the solution where the particles are embedded. The resonance frequency can thus be tuned by changing the trap stiffnesses and liquid viscosity, and increases with the (square root of) product of the two stiffnesses, while reducing with increasing viscosity. Thus, the amplitude and phase responses of the resonance can be used to determine the viscosity if the trap stiffnesses are known. We explicitly determine the analytical expressions for the sensitivity of the phase and amplitude responses with changing trap stiffness ratio as well as liquid viscosity, and observe that the responses are not very sensitive to trap stiffness ratio, but reduce with increasing liquid viscosity. We demonstrate theoretically and perform careful experiments to prove that the phase response has greater accuracy and precision in viscosity measurement compared to the amplitude by measuring the viscosity of water for different separations of the control and probe particles. We then proceed to determine the viscosity of different water and glycerol mixtures, and on comparing our results to that obtained on the same samples by a commerical rheometer, observe agreement to within 2\%. This is an improvement of around 2x in accuracy over our recently reported results in Ref. \citep{scirepbayes}, where we had demonstrated a one-point passive measurement of viscosity by a Bayesian analysis of the Brownian motion trajectories of a trapped particle. This method can be extended to n-particle active microrheology measurements in viscous fluids by holographic tweezers where different sets of control and probe particles can be used and the data collected simultaneously using fast cameras. More interestingly, the resonance characteristics of the probe in a viscoelastic medium should provide even more fascinating results, given that the hydrodynamic interactions in these cases are retarded. This should set up a new paradigm in the active microrheology of complex fluids using optical tweezers - the present set of experiments can be thought of as forming a baseline. We have commenced work in these areas and should be able to report new results in the near future.

\section*{Acknowledgments}

This work was supported by the Indian Institute of Science Education
and Research, Kolkata, an autonomous research and teaching institute
funded by the Ministry of Human Resource Development, Govt. of India.
We acknowledge Dr. Ronojoy Adhikari, Mr. Abhrajit Laskar, and Mr.
Rajesh Singh of The Institute of Mathematical Sciences for help in
developing the theoretical formalism. The authors also acknowledge
Mr. Chayan Dey and Dr. Prasun Mondal of Department of Chemical Sciences,
IISER Kolkata for their help in the rheometer measurements.

\bibliographystyle{apsrev4-1}
%

\end{document}